\documentclass[sigconf]{acmart}

\usepackage{makecell}
\AtBeginDocument{%
  \providecommand\BibTeX{{%
    \normalfont B\kern-0.5em{\scshape i\kern-0.25em b}\kern-0.8em\TeX}}}

\setcopyright{rightsretained}
\acmConference[SIGIR eCom'21]{ACM SIGIR Workshop on eCommerce}{July 15, 2021}{Virtual Event, Montreal, Canada}
\acmYear{2021}
\copyrightyear{2021}
\makeatletter
\renewcommand\@formatdoi[1]{\ignorespaces}
\makeatother
\acmISBN{}

\begin{document}

%% EXAMPLE DOC https://drive.google.com/file/d/1_ve441YsLxqSqArIE31gDWPl4pthAG5i/view?usp=sharing

\title{SIGIR 2021 E-Commerce Workshop Data Challenge}

%%
%% SORRY, I PUT COVEO PEOPLE FIRST AND THEN ALPHABETICAL
%% AS WE WILL PROVIDE THE DATA
%% HOPE IT'S OK

\author{Jacopo Tagliabue}
\affiliation{%
  \institution{Coveo Labs}
  \city{New York}
  \country{USA}
}
\email{jtagliabue@coveo.com}

\author{Ciro Greco}
\affiliation{%
  \institution{Coveo Labs}
  \city{New York}
  \country{USA}
}
\email{cgreco@coveo.com}

\author{Jean-Francis Roy}
\affiliation{%
  \institution{Coveo}
  \city{Quebec City}
  \country{Canada}
}
\email{jfroy@coveo.com}

\author{Bingqing Yu}
\affiliation{%
  \institution{Coveo}
  \city{Montreal}
  \country{Canada}
}
\email{cyu2@coveo.com }

\author{Patrick John Chia}
\affiliation{%
  \institution{Coveo}
  \city{Montreal}
  \country{Canada}
}
\email{pchia@coveo.com }

\author{Federico Bianchi}
\affiliation{%
  \institution{Bocconi University}
  \city{Milano}
  \country{Italy}
}
\email{f.bianchi@unibocconi.it}

\author{Giovanni Cassani}
\affiliation{%
  \institution{Tilburg University}
  \city{Tilburg}
  \country{Netherlands}
}
\email{g.cassani@tilburguniversity.edu }

\renewcommand{\shortauthors}{Tagliabue et al.}

%\begin{abstract}
%  BLAH BLAH.
%\end{abstract}

%\keywords{recommender systems, intent prediction, e-commerce}

\maketitle

\section{The Coveo Data Challenge}
\label{sec:intro}

The massive growth of e-commerce - which is estimated to turn into a $4.5$ trillion industry in 2021~\cite{ecomworld} - has fueled significant investments in areas such as Information Retrieval, NLP and recommendation systems. One of the principal goals of such investments has been to improve personalization in the customer journey. Optimal customer journeys in e-commerce are ideally supported by systems that transform the experience of each customer based on customer behavior and (a minimal set of) business rules~\cite{Tsagkias2020ChallengesAR}.

Although compelling progress has been made, interesting challenges remain in personalization. In particular, customer intentions can be different depending on the occasion and even change within the same session, depending on where users are in their customer journey. In this context, the feedback loop determined by behavioural signals - such as searches, clicks, add-to-cart, cart-abandonment - spans from hours to a few seconds~\cite{10.1145/3159652.3159714} and machine learning models need to adapt as fast as possible to the continuously changing nature of the customer journey.

The need for efficient procedures for personalization is even clearer if we consider the e-commerce landscape more broadly. Outside of giant digital retailers, the constraints of the problem are stricter, due to smaller user bases and the realization that most users are not frequently returning customers: in dozens of deployments across several verticals, we found that it is not uncommon that only <30\% of shoppers visit more than 2 times a year. Because of these constraints, the profit increase due to personalization strategies~\cite{Gartner2018} has been often disappointing. In a context where user intentions are constantly changing, customers are mostly non-recurring and bounce rates are high~\cite{SimilarWeb2019}, the competitive pressure towards real-time personalization requires machine learning models to work~\textit{as early as possible} in the customer journey and with~\textit{as little data} as possible. In this perspective, sessions emerge as the natural partition for building models with strong business impact~\cite{Hidasi2016SessionbasedRW,STAMP,10.1145/3289600.3290975,10.1145/3357384.3357895,10.1145/3394486.3403278} and session-based inference a crucial topic for retailers ~\textit{of all sizes}~\cite{CoveoECNLP202,Bianchi2020FantasticEA}

For practitioners reading about newer models, it is often hard to understand how those advances translate into their production context (if at all~\cite{10.1145/3298689.3347041}), as researchers either use old (e.g. \textit{MovieLens}~\cite{10.1145/2827872} in~\cite{10.1145/3394486.3403278,schnabel2020debiasing,krichene2020sampled}), industry-specific datasets (e.g. playlists~\cite{10.1145/3240323.3240377}), or private datasets that may not be representative of common constraints~\cite{Toth2017PredictingSB}. For these reasons, we designed the~\textbf{Coveo Data Challenge} with two principles in mind: 

\begin{itemize}
    \item advancing the field with the release of a new, fine-grained, session-based dataset. Our dataset (Section~\ref{sec:data}) presents product-level data from a mid-sized e-commerce, coupled with search-based interactions and content-based information; as such, it is a representative example of thousands of shops in the middle of the industry long tail, with realistic traffic, conversion rate, abandonment ratio;
    \item proposing inference tasks that are interesting for retailers of all sizes, but where models are evaluated on a mid-size shop. Given the increasing commoditization of tools and libraries, we expect more shops to develop ML capabilities in the coming years: as such, we believe in the value of research with tangible return on investment even outside of retail giants. 
\end{itemize}

The rest of~\textit{this} document describes the data release, the chosen tasks and the evaluation procedure. Moreover, a brief survey of existing work is provided in Section~\ref{sec:related}.

\subsection{About the organizers}
The~\textbf{Coveo Data Challenge} is organized by a collaboration between academic researchers and~\textit{Coveo}, a multi-tenant provider of A.I. services, with a network of more than 1000 deployments for customer service, e-commerce and enterprise search use cases.

\section{Task Description}
\label{sec:problem}

For this data challenge, we release a new session-level dataset, containing 10M product interactions over 57K different products (Section~\ref{sec:data}). Starting from this dataset, we accept solutions addressing two different tasks:

\begin{enumerate}
    \item \textbf{a session-based recommendation task}, where a model is asked to predict the next interactions between shoppers and products, based on the previous product interactions and search queries within the session;
    \item \textbf{a cart-abandonment task}, where, given a session containing an add-to-cart event for a product~\textit{X}, a model is asked to predict whether the shopper will buy~\textit{X} or not in that very session. In particular, we are interested in modelling how abandonment prediction changes $1, 2, …, n$ events after the original event, with the idea that the biggest business value comes from early prediction.
\end{enumerate}

\section{Data Description}
\label{sec:data}

The~\textbf{Coveo Data Challenge} features the release of a dataset comprising three main files for training:

\begin{itemize}
    \item \begin{sloppypar}\textbf{browsing interactions}: a text file containing fine-grained shopping tracking for more than 4M sessions\footnote{The definition of ``session'' is the industry standard from~\url{https://support.google.com/analytics/answer/2731565?hl=en}.}. Each row is a separate interaction (page view or product interaction); events with the same~\texttt{session\_id\_hash} are from the same session -- please note that a product page may generate both a~\textit{detail} and a~\textit{pageview} event, and that the order of the events in the text file is not strictly chronological (refer to the session and timestamp information to reconstruct the actual chain of events for a given session). The fields are similar to the dataset in~\cite{Requena2020}\footnote{The dataset is available at: \url{https://github.com/coveooss/shopper-intent-prediction-nature-2020}.}:~\texttt{session\_id\_hash}, 
    \texttt{event\_type},  \texttt{product\_action}\footnote{Please note that removing a product from the cart might lead to multiple \textit{remove} events, hence pre-processing might be needed with regard to this type of interactions.}, \texttt{product\_sku\_hash}, \texttt{server\_timestamp\_epoch\_ms}\footnote{As a further anonymization technique, the timestamp has been shifted by an unspecified amount of weeks, keeping intact the intra-week patterns.}, \texttt{hashed\_url}. Descriptive statistics for this file can be found in Table~\ref{tab:freq};\end{sloppypar}
    \item \begin{sloppypar}\textbf{search interactions}: a text file containing more than 800k search-based interactions. Each row is a search query event issued by a shopper, which includes an array of (hashed) results returned to the client. We also provide which result(s) have been clicked from the result set, if any. By also reporting products seen but not clicked, we hope to inspire clever ways to use negative feedback. The fields are: \texttt{session\_id\_hash}, \texttt{server\_timestamp\_epoch\_ms}, \texttt{query\_vector}, \texttt{product\_skus\_hash},
    \texttt{clicked\_skus\_hash}. The field~\texttt{query\_vector} is a vectorized representation of the original search query, obtained through standard pre-trained models and dimensionality reduction techniques;\end{sloppypar}
    \item \begin{sloppypar}\textbf{product meta-data}: a text file containing a mapping between the SKUs (hashed) present in the entire dataset (training+testing) and their respective meta-data (when available), including a hashed representation of product categories, pricing information and vectorized representations of textual and image meta-data. The fields are:
    \texttt{product\_sku\_hash}, 
    \texttt{category\_hash},~\texttt{price\_bucket}, 
    \texttt{description\_vector}, 
    \texttt{image\_vector}. The categories are hashed representations of a category tree\footnote{The product categorization is given by the shop -- in this case, product experts have defined a custom hierarchy to classify products for navigation and marketing purposes.}, where each level of hierarchy is separated with a \texttt{/}. The pricing information is provided as a 10-quantile integer. The vectorized fields are a dense representation of the original catalog, obtained through standard pre-trained modeling and dimensionality reduction techniques.\end{sloppypar}
\end{itemize}

All files are available for download, and additional important information and sample scripts are filed under the project repository\footnote{The data challenge is hosted at~\textbf{https://github.com/coveooss/SIGIR-ecom-data-challenge}.}. All training data comes from sampling in an unspecified ratio\footnote{Exact traffic information is thus removed as a further anonymization strategy.} several months of interactions in the life of~\textit{Shop Z}, a mid-size shop with~\textit{Alexa Ranking} between 25k and 200k; test data is sampled from a disjoint and adjacent time period. Data is fully anonymized and hashed, and collected in compliance with existing legislation: download and usage of the data implies acceptance of the terms and conditions (Section~\ref{sec:legal}).

\begin{table}
  \caption{Descriptive statistics for the training dataset containing browsing interactions.}
  \begin{tabular}{lc}
    \toprule
    Property&Value\\
    \midrule
    \# of SKUs & $57,483$\\
    \# of sessions & $4,934,699$\\
    \# of all events & $36,079,307$\\
    \# of all product interactions & $10,431,611$\\
    \# of add-to-cart  & $329,557$\\
    \# of purchases & $77,848$\\
    25/50/75 pct. of session length& $2/3/8$\\
  \bottomrule
\end{tabular}
  \label{tab:freq}
\end{table}

\subsection{Comparison with existing datasets}
\label{sec:existing}
Based on recent recommendation surveys and frameworks~\cite{zhao2020recbole,ludewig2018evaluation} 
%(e.g. \cite{zhao2020recbole} that compares 53 models on 27 benchmark datasets), 
we can classify benchmarks in different categories including \emph{ratings} (e.g. MovieLens~\cite{10.1145/2827872}, Anime~\cite{kaggleanime}, Netflix~\cite{netflix}, Book-Crossing~\cite{ziegler2005improving}), \emph{reviews} (e.g. Epinions~\cite{zhao2014leveraging}, Yelp~\cite{yelp}, Steam~\cite{kang2018self}, Amazon~\cite{mcauley2015image}), \emph{advertising} (Criteo~\cite{criteo}, Avazu~\cite{avazu}) and \emph{e-commerce}. E-commerce datasets differ in terms of type of available data. Some only provide transactions (Ta-Feng, Belgium Retail Market, Foodarkt~\cite{recsys_grocery}) or product data for product classification~\cite{rakuten,kaggle_innerwear,kaggle_product_item}, whereas others provide also interactions and product metadata (YOOCHOOSE (RCS15)~\cite{yoochoose}, Retailrocket (RETAILR)~\cite{retailrocket}, DIGINETICA~\cite{diginetica}). Table~\ref{tab:data_comparison} compares these three datasets with the data distributed for this challenge. 

\begin{table}
  \caption{Comparison of e-commerce datasets from the literature with the Coveo dataset.}
  \label{tab:data_comparison}
  \begin{tabular}{lcccc}
    \toprule
      & Coveo & DIGINETICA & RETAILR & RCS15\\
    \midrule
    Searches & \checkmark & \checkmark & & \\
    Impressions & \checkmark & & & \\
    Clicks/Views & \checkmark & \checkmark & \checkmark & \checkmark \\
    Carts & \checkmark & & \checkmark &  \\
    Transactions & \checkmark & \checkmark & \checkmark & \checkmark \\
    \midrule
    Category &tree&single&tree&---\\[3mm]
    Price &\checkmark & \checkmark & --- & \checkmark\\[1mm]
    Description &embedded& \makecell{hashed\\name}& \makecell{hashed\\properties}& ---\\[3mm]
    Image &embedded&\checkmark&---&---\\[3mm]
    \midrule
    Users & --- & 88K & 1.4M & --- \\
    Items & 66K & 184K & 235K  & 52K \\
    Events & 37M & 3.3M & 2.7M & 42M \\
    \bottomrule
\end{tabular}
\end{table}

Considering the variety of the available features and the fine-grained details included in the~\textbf{Coveo Data Challenge}, we believe it to be a significant addition to the set of resources in the space. In particular, by releasing neural-friendly embeddings for text, images~\textit{and} search queries, we hope to strike a virtuous trade-off between anonymity and information content.

\section{Evaluation}
\label{sec:eval}

In this Section, we describe the evaluation procedure and the key metrics that will be used to evaluate submissions in our two tasks\footnote{The evaluation script can be found in the public repository. Please refer to the challenge website for latest news about important dates, rules and submission format.}. While the final ranking will solely reflect quantitative metrics, we encourage papers to include a section on qualitative evaluation, following the suggestions below.

\subsection{Quantitative evaluation}

\begin{enumerate}
    \item \textbf{Recommendation}. We follow the evaluation protocol described in~\cite{ludewig2018evaluation}: we reveal the start of a session to the model, and we ask it to predict future interactions. In particular, given a sequence of $n$ events as input, we use two evaluation schemes:
    \begin{itemize}
        \item the model is evaluated at predicting the \emph{immediate next item}, using the~\textbf{Mean Reciprocal Rank} (MRR) as our metric;  
        \item the model is evaluated at predicting \emph{all subsequent items} of the session, up to a maximum of 20 after the current event. We take the \textbf{F1 score} from precision@20 and recall@20 as the final metric.
    \end{itemize}
    Moreover, we also measure additional quality factors defined in~\cite{jannach2017recurrent,ludewig2018evaluation}: coverage@20 and popularity bias@20. Finally, note that when a test session includes a search query, this event will be available to the model as an~\emph{input} (it can and should be taken into consideration by the model), but it will be ignored when computing the metrics: in other words, models are evaluated only on their ability to predict future interactions with products.
    \item \textbf{Cart-abandonment}. Models will be evaluated by their \textbf{micro F1 score} in predicting shopping cart abandonment at the first add-to-cart event (AC), and then 2, 4, 6, 8 and 10 events after it.\footnote{By assessing performance at AC, moreover, we want to stress the importance of predicting intent from as little data as relevant.} Each model will be evaluated by performing a weighted combination of micro F1 scores at different clicks, with larger weights assigned to earlier predictions according to the following schema: 
    \begin{itemize}
        \item micro F1 at AC * 1
        \item micro F1 2 clicks after AC * 0.9
        \item micro F1 4 clicks after AC * 0.8
        \item micro F1 6 clicks after AC * 0.7
        \item micro F1 8 clicks after AC * 0.6
        \item micro F1 10 clicks after AC * 0.5
    \end{itemize}
    The test set will not undergo any resampling to even the class distribution. Test sessions including purchases will be trimmed before the first purchase event to avoid trivializing the inference. 
\end{enumerate}

\subsection{Qualitative evaluation}

While we \textit{do} recognize the importance of a standardized quantitative evaluation to have some measure of progress, we also agree with recent criticism of ``leaderboard chasing''~\cite{Ethayarajh2020UtilityII}, as there are many practical constraints not captured by metrics which are crucial for the real-world success of a model. 

To encourage a deeper understanding of the underlying business problems and foster a healthy competition among models with broad applicability, we solicit submissions that provide new insights by mixing quantitative results with industry-relevant qualitative discussion. As an inspiration, we provide below some examples of interesting topics, divided by task: as a general recommendation, reflections on training cost, inference time and sample efficiency are greatly appreciated, together with connections to related issues in IR for products~\cite{coveoNAACLINDUSTRY21}.

\begin{enumerate}
    \item \textbf{Recommendation}. We encourage papers to find solutions creatively using search and content data, together with browsing behavior. In particular, we encourage error analysis and comments on model performance for slices of the data that reflects practically important scenarios -- new items, rare items, after-query prediction, interplay between search behavior and browsing patterns.
    \item \textbf{Cart-abandonment}. We encourage submissions to include a careful error analysis which considers the relation between classification accuracy at different target clicks and different session characteristics, such as the number of interactions prior to the cart event and the total length of the session~\cite{Requena2020}. For example, suppose there are 15 actions after the AC in a purchase session: a model which converges on consistently predicting a purchase from action 11 (not captured in our evaluation) would anticipate the right outcome more than a model which is undecided until the second-to-last action but still correctly predicts a purchase at the end of the session. A measure of anticipation would be very informative although it will depend on some free parameters (how strong should the support for a given prediction be, for how many consecutive actions should it remain stable, etc.): are there important insights of this sort that we can uncover leveraging this new dataset?
\end{enumerate}

\section{Related Work}
\label{sec:related}

\textbf{Session-based models}. Session-based inference is a well-studied topic in e-commerce, as it powers recommendation~\cite{Wu2019SessionbasedRW,10.1145/3357384.3357964, Yu2020TAGNNTA,Zheng2020DGTNDG,Bianchi2020BERTGS} and NLP/IR use cases~\cite{10.1145/3366424.3386198,Tagliabue2020ShoppingIT}: while neural models are now dominating the research agenda~\cite{10.1145/3394486.3403278,10.1145/3357384.3357895}, there is also some skepticism about the widespread applicability of these techniques and the progress made by the field~\cite{10.1145/3298689.3347041,Ludewig2019EmpiricalAO}; in particular, a significant number of recent models still goes through sub-optimal benchmarks~\cite{10.1145/3394486.3403278,schnabel2020debiasing,krichene2020sampled}, which may not be representative of the type of constraints faced by many practitioners~\cite{10.1145/3415959.3416001}.

\textbf{Intent prediction}. The ability to predict user intent (add to cart, purchase, abandonment, etc.) from click-stream data is a long-standing challenge for e-commerce, with older solutions modelling user browsing in a Markovian fashion~\cite{doi:10.1137/1.9781611972733.10} and recent attempts leveraging deep learning architectures~\cite{Bigon2019PredictionIV}. As for in-session recommendations, the broad impact of recent  findings is unclear: for example, the architecture proposed in~\cite{Toth2017PredictingSB} was tested with an unrealistic conversion rate (10x a ``standard'' shop). More recently,~\cite{Requena2020} simulated accuracy in deployment scenarios by artificially sub-sampling raw data, showing that models outperform a simple baseline only up to a class imbalance of 80\% to 20\%. More research into models that scale to extreme class imbalance is thus a priority, and with our release we hope to nudge the community to devote more resources to this problem.

\section{Legal Notice}
\label{sec:legal}
The~\textbf{Coveo Data Challenge} comes with specific Terms and Conditions, as found in the public repository and bundled with the dataset archive: by downloading and accessing the data, you agree to comply with the terms -- in addition, by submitting to the challenge, you agree to release your code under an open source license. In particular, the dataset is freely provided to the research community with the purpose of advancing the field: all data was collected and processed in an anonymized fashion, in compliance with existing legislation, and released with additional hashing of meta-data: participants agree to not use the dataset for any other purpose than what is stated in the Terms and Conditions, nor attempt to reverse engineer or de-anonymise the dataset by explicitly or implicitly linking the data to any person, brand or legal entity.

\begin{acks}
The organizers wish to thank Luca Bigon for his outstanding support in data collection, Surya Kallumadi, Massimo Quadrana, Dietmar Jannach and Ajinkya Kale for the precious feedback on a previous version of this paper. Finally, special thanks to Coveo's legal team for believing in this data sharing initiative.
\end{acks}

\bibliographystyle{ACM-Reference-Format}
\bibliography{new}

\end{document}